\documentclass[superscriptaddress,secnumarabic,amssymb,amsmath,nobibnotes,aps,prd,nofootinbib,showkeys,preprint]{revtex4}

\usepackage{graphicx}
\usepackage{float}
\usepackage{bm}
\usepackage{amsmath}
\usepackage{amsfonts}
\usepackage{amssymb}
\usepackage{epstopdf}
\usepackage{natbib}%
\newcommand{\bee}{\begin{equation}}
\newcommand{\eee}{\end{equation}}
\newcommand{\eaa}{\end{eqnarray}}
\newcommand{\baa}{\begin{eqnarray}}

\usepackage{color}

\begin{document}
\title{Exploring modified Kaniadakis entropy: MOND-related theory, the Bekenstein bound conjecture, and Hawking evaporation 
within the Landauer principle}

\author{Gabriella V. Ambrósio}
\email{gabriellambrosio@gmail.com}
\affiliation{Departamento de F\'isica, Universidade Federal de Juiz de Fora, Juiz de Fora – 36036-330, MG, Brazil}
\author{Michelly S. Andrade}
\email{michelly.andrade32@gmail.com}
\affiliation{Departamento de F\'isica, Universidade Federal de Juiz de Fora, Juiz de Fora – 36036-330, MG, Brazil}
\author{Paulo R. F. Alves}
\email{paulo.alves@ice.ufjf.br}
\affiliation{Departamento de F\'isica, Universidade Federal de Juiz de Fora, Juiz de Fora – 36036-330, MG, Brazil}
\author{Cleber N. Costa}
\email{cleber.costa@ice.ufjf.br}
\author{Jorge Ananias Neto}
\email{jorge.ananias@ufjf.br}
\affiliation{Departamento de F\'isica, Universidade Federal de Juiz de Fora, Juiz de Fora – 36036-330, MG, Brazil}
\author{Ronaldo Thibes}
\email{thibes@uesb.edu.br}
\affiliation{Universidade Estadual do Sudoeste da Bahia, DCEN,
Rodovia BR 415, Km 03, Itapetinga – 45700-000, BA, Brazil}

\begin{abstract}
We investigate the description of black-hole thermodynamics in terms of a recently proposed modified version for Kaniadakis entropy.
We discuss the role of that proposal within the Modified Newtonian 
Dynamics (MOND) theory, a generalization of Newton's second law aimed at explaining galaxy rotation curves without resorting to dark matter.
We posit 
a conjecture that the Kaniadakis entropy precisely describes the Bekenstein-Hawking black-hole entropy.  
Furthermore, we consider the Bekenstein bound conjecture which imposes an upper limit on the entropy of confined quantum systems. 
We analyze that conjecture in the context of the modified Kaniadakis entropy and find that it holds for typical values of $\kappa$, 
as evidenced by our numerical investigation.
Finally, using the Landauer principle from information theory, we derive an expression for mass loss in black hole evaporation.
Our exploration underscores the potential relevance of a modified Kaniadakis statistics in understanding diverse physical phenomena, 
from gravitational systems to quantum mechanics, offering a promising direction for future research at the intersection among statistical 
mechanics and a continually increasing number of other important areas of physics.
\end{abstract}


\maketitle
\section{Introduction}

Hawking's discovery \cite{swh} of thermal radiation emanating from a black-hole (BH) came as a surprise to most experts, 
despite some prior indications of a 
significant link between BH physics and thermodynamics. Bekenstein \cite{jdb} observed that certain properties of BHs, such as 
their area, bear resemblances 
to entropy. Indeed, according to the Hawking area theorem \cite{swh}, the BH area $A$ does not decrease in any classical 
scenario, behaving very similarly 
to entropy. Since then, it has been clear that the analogies between BH physics and thermodynamics are not a mere coincidence 
and quite a broad relation 
connecting those two fields has been established. Just as any typical thermodynamical system, an arbitrary BH heads towards 
equilibrium, achieving a steady 
state after a relaxation process. Not long ago, the implications of introducing a fractal structure to the horizon geometry of 
BHs have been investigated in references \cite{barrow, aab,aa1,aa2}.

Another intriguing phenomenon we shall discuss here concerns the behavior of galaxy rotation curves, an observational fact for 
which two main distinct 
contending models can be found in the literature: the first one relies on the introduction of dark matter to accommodate for 
the observed data, while the 
second one attempts to modify Newton's fundamental law of dynamics. 
In the present article, we focus on the latter one, referred in the literature by the acronym MOND for Modified 
Newtonian Dynamics \cite{mm1,mm2,mm3}. It is important to mention though that, as a fundamental theory, MOND certainly possesses 
limitations. In fact, the universality of its proposed relation at galactic scales has been recently challenged in the literature, suggesting that 
the constant $a_0$ (see Eq. (\ref{mond}) below) may not enjoy true universality, as evidenced by the absence of a fundamental acceleration scale 
in galaxies \cite{Rodrigues:2018duc}. Relativistic extensions of MOND, such as the TeVeS model, encounter significant difficulties in accurately 
reproducing the cosmic microwave background (CMB) spectrum \cite{Xu:2014doa}.
Connecting the previous subjects, we explore BH physics via MOND within the framework of a proposed modification for Kaniadakis entropy \cite{aa}.

Interestingly, the Landauer principle \cite{landauer} states that erasing one bit of information in a computational system requires 
a minimum amount 
of energy, resulting in heat dissipation. This energy is proportional to the temperature of the system and is fundamental to the 
thermodynamics of information processing. It points out the physical limits of computation, linking information theory with thermodynamics.

The Kaniadakis entropy belongs to a class of non-Gaussian entropies which have received significant attention in BH thermodynamics, 
as evidenced by 
the extensive research cited in references \cite{ci,tc,bc,mzlgsj}. Notably, the authors of references \cite{ci,bc} have introduced a 
novel variant of 
Rényi entropy for BH horizons. That is achieved by treating BH entropy as a nonextensive Tsallis entropy and employing a logarithmic formula. 
This approach yields a dual Tsallis entropy whose nonextensive effects enable the stabilization of BHs.
Motivated by that result, one of the authors of the present work has recently proposed a dual Kaniadakis entropy \cite{aa}, 
where the stabilization of BHs has also been achieved.

The purpose of this paper is to utilize Kaniadakis' modified entropy in three situations. In the first one,
a consistent modification for the gravitational force will be derived from the concept of entropic force, where the notion 
of a holographic surface 
entropy will prove to be crucial. As we shall examine, this derived effective gravitational force exhibits certain parallels to the MOND theory.
In the second one, we shall demonstrate that Kaniadakis' modified entropy satisfies the Bekenstein bound conjecture \cite{beke}, 
an important proposition 
establishing an upper limit for entropies. 
In the third case, within the context of information theory, specifically Landauer's 
theory, we will determine the mass loss of the black hole corresponding to a one-bit reduction in information.

\section{Kaniadakis statistics}

Kaniadakis statistics, also known as $\kappa$-statistics \cite{k1,k2,k3,k4}, offers a non-extensive generalization for the standard Boltzmann-Gibbs (BG) 
statistics. Similar to Tsallis thermostatistics, it introduces a parameter $\kappa$ modifying the usual entropy definition. More precisely, the $\kappa$-entropy is defined as
\begin{eqnarray}
S_\kappa = - k_B  \sum_i^W \, \frac{p_i^{1+\kappa} - p_i^{1-\kappa}}{2 \kappa} \,,
\end{eqnarray}
where $k_B$  is the Boltzmann constant, $p_i$ are the probabilities associated with each microstate, $\kappa$ is a real parameter and $W$ is the total 
number of microstates.
This equation recovers the Boltzmann-Gibbs (BG) entropy when $\kappa$ approaches 0. Remarkably, $\kappa$-entropy exhibits the main 
properties of an entropy except for additivity. It is actually found that $\kappa$-entropy satisfies a {\it pseudo-additivity} property.
An important aspect of $\kappa$-entropy is its interpretation as a relativistic generalization of BG entropy. This framework has proven 
successful in various experimental contexts, including cosmic rays \cite{ks,kqs}, cosmic effects \cite{aabn} and gravitational 
systems \cite{aamp, lbs, ggl,as}.
Within the microcanonical ensemble, where all states have equal probability, Kaniadakis entropy simplifies to

\begin{eqnarray}
\label{km}
S_\kappa = k_B\,\frac{W^\kappa - W^{- \kappa}} {2 \kappa} \,.
\end{eqnarray}
This relation reproduces the standard BG entropy formula $ S = k_B \,\ln W$ in the limit $\kappa \rightarrow 0$.

In a previous paper \cite{aa}, we have proposed that the Kaniadakis entropy, Eq. (\ref{km}), can describe the BH entropy, $S_{BH}$, 
from the equality
\begin{eqnarray}
\label{kbh} 
k_B\, \frac{W^\kappa - W^{- \kappa}} {2 \kappa} = S_{BH}\,.
\end{eqnarray}
Therefore, from Eq. (\ref{kbh}), we have
\begin{eqnarray}
\label{w}
W = \left(  \kappa \, \frac{S_{BH}}{k_B} + \sqrt{ 1+\kappa^2 \frac{S_{BH}^2}{k_B^2} } \right)^\frac{1}{\kappa} \,.
\end{eqnarray}
Using Eq. (\ref{w}) with the BG entropy, we obtain
\begin{eqnarray}
\label{kmod}
S^*_\kappa = \frac{k_B}{\kappa}  \, \ln \left(   \kappa \, \frac{S_{BH}}{k_B} + \sqrt{ 1+\kappa^2 \frac{S_{BH}^2}{k_B^2} }     \;           \right) \,.     
\end{eqnarray}
Thus, Eq. (\ref{kmod}) represents a deformed version for the Kaniadakis entropy, which we denote here as the 
{\it modified Kaniadakis entropy}. Note that, when we assume $\kappa = 0$ in Eq. (\ref{kmod}), $S_\kappa^*$ becomes $S_{BH}$. 
We can express the entropy $S_{BH}$ in (\ref{kmod}) as
\begin{eqnarray}
\label{sbh}
S_{BH} = \frac{k_B c^3 A}{4 G \hbar}     \,,     
\end{eqnarray}
where $c$ is the speed of light, $G$ is the gravitational constant, $\hbar$ is the Planck constant and $A$ is the area of the event horizon.
In terms of the Planck length $l_p$ and the radius of the event horizon $R$, where we use $A = 4 \pi R^2$, the entropy $S_{BH}$ can be rewritten as
\begin{eqnarray}
\label{sbhr}
S_{BH} = k_B \, \frac{\pi R^2}{l_p^2} \,.     
\end{eqnarray}
Here it is important to mention that the Hawking temperature obtained through the modified Kaniadakis entropy, Eq. (\ref{kmod}), 
leads to the Schwarzschild BH heat capacity being positive, which means that the Schwarzschild BH described by the modified Kaniadakis 
entropy is thermodynamically stable. For more details, see reference \cite{aa}.

\section{MOND-related theory}

The MOND theory success stems from its capacity, as a phenomenological model, to effectively elucidate the majority of galaxies' rotation 
curves. It aptly reproduces the renowned Tully-Fisher relation \cite{tf}  and offers a feasible alternative to the dark matter 
model. Nevertheless, it is worth noting that MOND theory falls short in explaining the temperature profile of galaxy clusters 
and encounters challenges when confronted with cosmological phenomena -- see references \cite{anep, cs, eumond} for a comprehensive 
discussion. Fundamentally, this theory represents a modification of Newton’s second law, wherein the force can be expressed as 
\begin{eqnarray}
\label{mond}
F = m \mu\left(\frac{a}{a_0}\right) \, a \,,
\end{eqnarray}
with $a$ denoting the usual acceleration, $a_0$ a suitable constant and $\mu(x)$ a real function possessing certain characteristics
\begin{eqnarray}
\label{mu1}
\mu(x)\approx  1 \;\; for \;\;  x >> 1 \,,
\end{eqnarray}
and
\begin{eqnarray}
\label{mu2}
 \mu(x) \approx x \;\; for \;\; x << 1\,.
\end{eqnarray}
Several proposals for $\mu(x)$ can be found in the literature \cite{fgb,zf}, yet it is commonly accepted that the core implications of 
MOND theory remain independent of the specific form of these functions.

With the aim of deriving a specific form for $\mu(x)$ from entropic principles, we start with a general expression for the force, 
governed by the equation of thermodynamics, given by \cite{mondnos,modran}
\begin{eqnarray}
\label{nl}
 F =  \frac{G M m}{R^2} \, \frac{4l_p^2}{k_B} \,\frac{dS}{dA} \,,
\end{eqnarray}
where $A$ is the area of the holographic screen and $S$ is the entropy describing it.
Note that expression (\ref{nl}) is general enough to allow for the consideration of any type of entropy. For example,
if we consider the Bekenstein-Hawking entropy law, $ S_{BH} = k_B \,\frac{A}{4l_p^2}$,
as the entropy $S$ in Eq. (\ref{nl}), then                        
we can recover the usual Newton's universal law $ F = G M m/R^2$.

It is important to mention here the work by Obregón~\cite{obregon}, where the gravitational force is modified using a generalized 
entropy based on superstatistics and the holographic principle. Also noteworthy is the systematic analysis by Martínez-Merino, 
Obregón, and Ryan~\cite{mor}, which examines various entropy formulations and their implications for the gravitational force, potential, 
and effective temperature.

So, in order to derive a new effective gravitational force law in the context of Kaniadakis modified entropy, initially we have
\begin{eqnarray}
\label{dsa}
\frac{dS_\kappa^*}{d A} = \frac{  \frac{  dS_{BH}  }{   dA }   }{  \sqrt{ 1 + \kappa^2 \frac{S_{BH}^2}{k_B^2}    }      } \,.
\end{eqnarray} 
Combining  Eqs. (\ref{sbhr}), (\ref{nl}) and (\ref{dsa}), we obtain a new effective gravitational force law as
 \begin{eqnarray}
 \label{fmond}
 F_{effective} = \frac{ \frac{G M m}{R^2}  } { \sqrt{ 1 + \left(   \frac{\kappa \pi R^2}{ l_p^2 }   \right)^2 } } \,.
\end{eqnarray}
We note that, contrary to the standard MOND theory where the decay is of the form $1/R$, Eq. (\ref{fmond}) decays with $1/R^4$ for large $R$.
After some algebra, we achieve
\begin{eqnarray}
 \label{fmond2}
 F_{effective} = \frac{ m \, a_N}{  \sqrt{ 1 + ( \frac{ a_0}{  a_N}    )^2 } } \,,
\end{eqnarray}
with 
\begin{eqnarray}
 \label{ac}
 a_N \equiv \frac{G M}{R^2}                   \,,
\end{eqnarray}
and

\begin{eqnarray}
 \label{a0}
a_0 \equiv  \frac{\kappa \pi G M}{l_p^2} \,.
\end{eqnarray}
Thus, from the effective gravitational force, Eq. (\ref{fmond2}), we can identify
\begin{eqnarray}
 \label{inter}
 \mu \left( \frac{    a_N     }{a_0} \right) = \sqrt{ \frac{1} { 1 +  \left( \frac{a_0}{   a_N } \right) ^2} } = \frac{    a_N  }{a_0} \left( 1+ \frac{   a_N ^2}{a_0^2} \right)^{- 1/2}   \,.
\end{eqnarray}
The interpolating function, Eq.(\ref{inter}), clearly satisfies conditions (\ref{mu1}) and (\ref{mu2}).
Therefore, we have shown that, by using Kaniadakis' modified entropy, Eq. (\ref{kmod}), along with the force expression in Eq. (\ref{nl}), 
we obtain a model similar to the MOND theory
with the corresponding interpolating function  Eq. (\ref{inter}).
It is intriguing to note that this interpolating function is precisely the standard interpolating function of 
MOND theory.

\section{Bekenstein bound conjecture}

As is well-known, the entropy of black holes is indeed proportional to the area of their event horizon. This principle, commonly 
referred to as the Bekenstein-Hawking entropy, stands as a cornerstone in BH thermodynamics. It highlights the intrinsic relationship 
between the BH entropy and the surface area of its event horizon. This concept, pioneered by Jacob Bekenstein and further elucidated 
by Stephen Hawking, holds significant importance in our comprehension of the thermodynamic properties of black holes and their 
implications for fundamental physics.

Additionally, Bekenstein proposed a universal upper limit on the entropy of a confined quantum system \cite{beke}, formulated as
\bee
\label{bekenstein1}
S \leq \frac{2 \,\pi\, k_B\, R\, E}{\hbar \,c} \,,
\eee
where $S$ represents entropy, $E$ is the total energy, and $R$ is the radius of a sphere containing the system. This inequality, 
known as the Bekenstein bound conjecture, implies that, as the Planck constant $\hbar$ approaches zero, the entropy becomes unbounded, 
indicating that the entropy of a localized system is unrestricted in a classical regime. Remarkably, this formulation excludes 
Newton's gravitational constant $G$, suggesting that the Bekenstein bound conjecture may extend beyond gravitational phenomena.

Despite some existing counter-examples suggesting possible flaws \cite{unruh,page,unruh2,gxo}, numerous instances  affirm the validity 
of the Bekenstein bound conjecture \cite{beke2,beke3, beke5, beke4}, rigorously demonstrated using conventional quantum mechanics and 
quantum field theory in flat spacetime \cite{cas}. Notably, the generalized uncertainty principle also contributes to deriving the 
Bekenstein bound \cite{bg}. The Bekenstein bound conjecture has a wide range of applications, including in cosmology and quantum field 
theory. See references \cite{bf,fs} for more details.

Moving forward, we next adopt the natural units system in which
\bee
\label{uni}
k_B = \hbar = c = G = 1  \,.
\eee
For these values, Eq. (\ref{bekenstein1}) simplifies to
\bee
\label{bekenstein2}
S \leq 2 \,\pi\, R\, E \,.
\eee
To derive a form compatible with BH physics, we start with the Schwarzschild metric given by
\bee
\label{sch}
ds^2 = \left( 1 - \frac{2 M}{R} \right) dt^2 - \left( 1 - \frac{2 M}{R} \right)^{-1} dr^2 - r^2 d \Omega^2 \,.
\eee
The solution for black holes leads to the singularity
\bee
\label{bhs}
R = 2 M \,.
\eee 
Assuming the radius $R$ of the enclosing sphere corresponds to the Schwarzschild radius and considering
$E=M$, Eq. (\ref{bekenstein2}) can be rewritten as
\bee
\label{bekenstein3}
S \leq \pi R^2 \,.
\eee
We observe that Eq. (\ref{bekenstein3}) reaches its maximum when $ S = \pi R^2 $ for a Schwarzschild BH, with entropy given by $ S_{BH} = A_H/4 $ where $ A_H $ represents the horizon area, equal to $ 4 \pi R^2 $.

Hence, considering that the modified Kaniadakis entropy governs BH thermodynamics, Eq. (\ref{kmod}) with $S_{BH} = \pi R^2$
allows us to recast the Bekenstein bound in
Eq. (\ref{bekenstein3}) as
\begin{eqnarray}
\label{kbb2}
S^*_\kappa \leq \frac{e^{\kappa S^*_\kappa} -   e^{- \kappa S^*_\kappa}    }{2\kappa}  \,,  
\end{eqnarray}
which can also be written in the form

\begin{eqnarray}
\label{kbb2h}
S^*_\kappa \leq  \frac{\sinh \kappa S^*_\kappa  } {\kappa } \,.
\end{eqnarray}
Thus,
Eq. (\ref{kbb2h}) represents the Bekenstein bound conjecture elaborated in the current context of modified Kaniadakis entropy.  In order to observe the veracity of
the inequality (\ref{kbb2h}), we have plotted in Fig. 1 the ratio
\begin{eqnarray}
\label{kbb3}
R_\kappa \equiv  \frac{\sinh \kappa S^*_\kappa  }{\kappa S^*_\kappa}  \,,     
\end{eqnarray}
for typical values $\kappa \geq 0$ and $S^*_\kappa = 2$.
\begin{figure}[H]
	\centering
	\includegraphics[height=7 cm,width=10 cm]{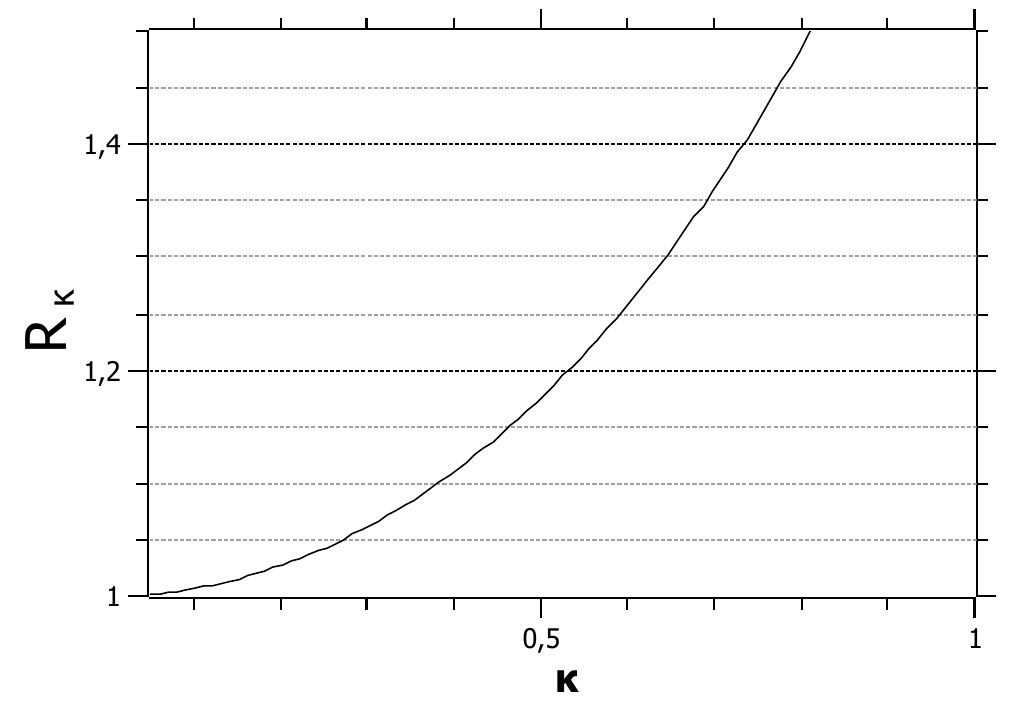}
	\caption{Values of $  R_\kappa \equiv  \frac{\sinh \kappa S^*_\kappa  }{\kappa S^*_\kappa} $ 
	as a function of $\kappa$ for $ S^*_\kappa = 2$.}
	\label{kaniadakisb}
\end{figure}
In Figure 1, two notable findings stand out. First, when $\kappa$ approaches 0, we observe that $R_\kappa$ goes to 1. This outcome 
can be analytically confirmed by employing power series expansions of hyperbolic sine functions for $\kappa$ values, resulting in 
the following approximation: $\sinh \kappa S^*_\kappa\approx  \kappa \, S^*_\kappa$. Consequently, it follows that $ R_\kappa $ 
equals 1, thus affirming the equality stated in Eq. (\ref{kbb2h}). The second result is that when the $\kappa$ parameter increases, the 
value of $R_\kappa$, which is always greater than one, increases. This result shows that the Bekenstein bound conjecture is satisfied 
when the modified Kaniadakis entropy is used to describe the black holes thermodynamics. 
Here it is important to mention that one of the authors of this work has shown that the usual Tsallis and Kaniadakis entropies, 
as well as the Barrow entropy, do not respect the Bekenstein bound conjecture. For further details, see reference \cite{euev}.

\section{Landauer principle}

The Landauer principle\footnote{Also referred as the Brillouin principle \cite{bri}.}, proposed in 1961 by Rolf Landauer \cite{landauer}, 
represents a fundamental concept that bridges information theory and thermodynamics.
It states that erasing a single bit of information incurs a minimum energy cost given by the equation
\begin{eqnarray}
 \label{land}
 \Delta E \geq k_B T \ln 2 \,,
\end{eqnarray}
where $k_B$ denotes Boltzmann's constant and $T$ the absolute temperature.
The $\ln 2$ term in Eq. (\ref{land}) arises from the fact that, considering the entropy of an information system (a measure of the 
system disorder), the number of possible microstates (configurations) is given by $\Omega = 2^N $, where $N$ is the number of bits. 
The energy defined in (\ref{land}) is dissipated as heat into the surroundings.
The Landauer principle brings attention to the physical nature of information and its inherent connection to thermodynamics. 
Any operation that manipulates information is therefore subject to thermodynamic principles. Logical irreversibility, or 
computationally irreversible processes, leads to physical irreversibility, which brings about an increase in entropy. Consequently, 
the act of erasing information results in heat dissipation, in accordance with the second law of thermodynamics, which states 
that entropy in an isolated system cannot decrease.
For instance, to delete a bit of data from a computer's memory, the physical state of the memory cell must be reset, requiring a 
certain amount of energy. This energy is released as heat, contributing to the computer overall energy consumption.

Although there have been some claims challenging its validity \cite{norton,norton2}, Landauer’s principle has consistently withstood 
extensive scrutiny over time.
Some experiments \cite{bap,jgb,yxr} have consistently demonstrated the energy cost associated with erasing information, aligning closely 
with the theoretical predictions. Additionally, rigorous mathematical proofs \cite{rw,ebor,ebor2} have further reinforced the Landauer
principle, establishing its important role in the field of thermodynamics of computation. 

In the recent article \cite{cl}, Landauer's principle has been applied to black hole physics in the context of Hawking evaporation.  The authors of \cite{cl} have noted that Hawking evaporation precisely saturates Landauer's principle, with the information lost by the black hole
during its evaporation occurring in the most efficient possible manner.  Following that, two late additional works \cite{otr, bgs} 
concerning applications of Laundauer's principle have just emerged: the first one to the apparent cosmological horizon \cite{otr} and the 
second one to the quantization of the black hole area \cite{bgs}.  It is important to note that, previously, Abreu had already linked 
Hawking temperature with Landauer's principle in \cite{abreu}.

In this context, it is then possible to derive Landauer's principle from the Bekenstein-Hawking entropy law. To begin with, 
proceeding with the natural units system in which $\,\hbar = c = G = 1$, we recall that 
the entropy of a black hole in terms of its mass is given by
\begin{eqnarray} 
\label{bhe} 
S = 4 \pi k_B  M^2 \,.
\end{eqnarray} 
From Eq. (\ref{bhe}), the temperature of the black hole is determined by
\begin{eqnarray} 
\label{bht} 
\frac{dS}{dM} = \frac{1}{T} = 8 \pi k_B  M \,. 
\end{eqnarray} 
Now, consider a scenario in which the black hole loses enough mass to reduce its information content by one bit. Using Eq. (\ref{bhe}), 
the mass loss results in
\begin{eqnarray} 
\label{dm} 
\Delta S = 8 \pi k_B  M \Delta M \,. 
\end{eqnarray} 
Assuming the change in entropy is
\begin{eqnarray} 
\label{dsb} 
\Delta S = k_B \ln 2 \,, 
\end{eqnarray} 
Eq. (\ref{dm}) can be rewritten as
\begin{eqnarray} 
\label{ldm} \Delta M = \frac{\ln 2}{8 \pi } \,\frac{1}{M}\,. 
\end{eqnarray} 
Substituting Eq. (\ref{bht}) into Eq. (\ref{ldm}) yields
\begin{eqnarray} 
\label{lds} 
\Delta M = k_B T \ln 2 \,. 
\end{eqnarray} 
Eq. (\ref{lds}) represents the Landauer principle in the case where it is saturated. Therefore, we can observe that the 
information loss to a black hole 
during its evaporation proceeds with maximum efficiency.

Our goal now is to calculate the mass loss of the black hole evaporation in the context of Kaniadakis modified entropy. To do this, 
we begin by  calculating the entropy variation, $\Delta S$, from Eq. (\ref{kmod}) where $S_{BH}$ is given by Eq.(\ref{bhe}), resulting 
in
\begin{eqnarray} 
\label{dmk} 
\Delta S = \frac{8 \pi  k_B  M \Delta M} { \sqrt{1+\kappa^2 (4 \pi  M^2)^2}} \,. 
\end{eqnarray} 
Using Eq. (\ref{dsb}) in (\ref{dmk}), we then obtain
\begin{eqnarray} 
\label{dmassk} 
\Delta M =  \frac{\ln 2}{8 \pi } \, \frac{ \sqrt{1+\kappa^2 (4 \pi  M^2)^2}}{M} \,. 
\end{eqnarray}
From Eq. (\ref{dmassk}), setting $\kappa = 0$, we can see that Eq. (\ref{ldm}) is then recovered. 
In Fig. 2, we have plotted the normalized mass variation, from Eq. (\ref{dmassk}), as a function of the mass $M$  
for $\kappa = 0$ (blue dash) and $\kappa = 0.2$ (red line).
\begin{figure}[H]
	\centering
	\includegraphics[height= 7 cm,width= 10 cm]{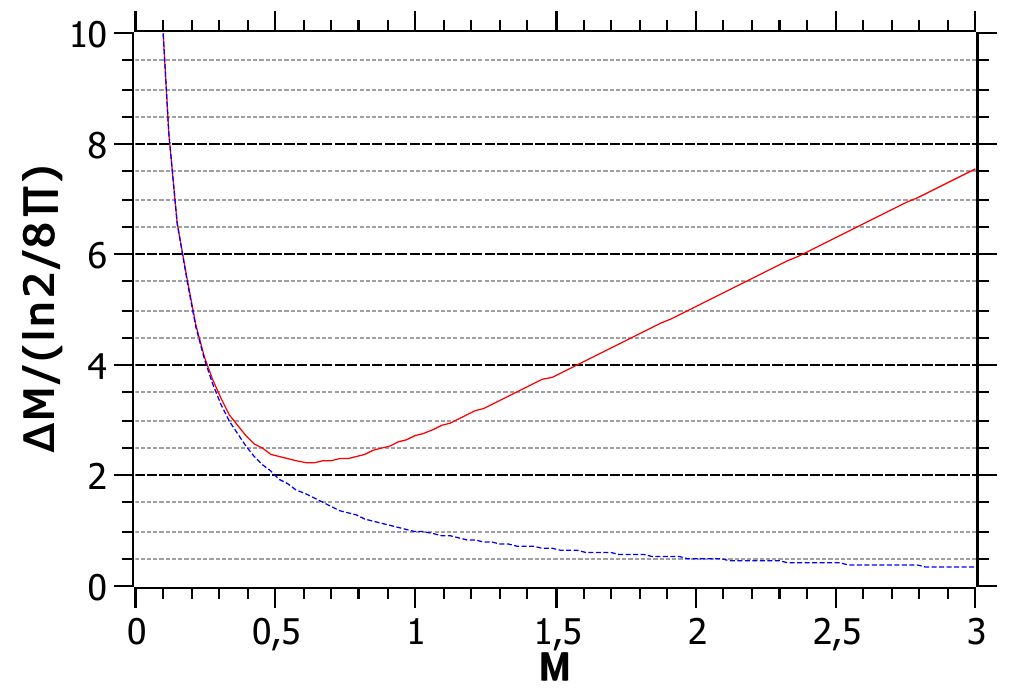}
	\caption{Values of the normalized black hole mass loss $  8 \pi {\Delta M}/{\ln 2} \, =\, 
 \frac{ \sqrt{1+\kappa^2 (4 \pi  M^2)^2}}{M}\,$ 
	as a function of $M$.}
	\label{landfig}
\end{figure}
From Fig. 2, we can observe a minimum value for the black hole mass loss $\Delta M$, Eq. (\ref{dmassk}), where $\kappa = 0.2$ (red line). 
From this minimum value, it can be observed that as $M$ increases, $\Delta M$  also increases, in contrast to what is noted in Eq. (32), 
where as $M$ increases, $\Delta M$ decreases.
This is an important result obtained by applying the modified Kaniadakis entropy to Landauer's principle, specifically in the context 
of black hole Hawking evaporation.

As a consequence of this result, we may comment on two aspects of the behavior illustrated in Fig.~2. 
The first aspect is that the black hole temperature in the modified Kaniadakis framework is defined by
\begin{equation}
T_\kappa = \left( \frac{dS^*_\kappa}{dM} \right)^{-1} \,,
\end{equation}
where \(S^*_\kappa\) is given in Eq.~(\ref{kmod}). Using Eq.~(\ref{bhe}), this yields
\begin{equation}
\label{tk}
T_\kappa = \frac{\sqrt{1+\kappa^2 (4 \pi M^2)^2}}{8\pi k_B M} \,.
\end{equation}
Combining Eq.~(\ref{tk}) with Eq.~(\ref{dmassk}) yields
\begin{equation}
\label{lpk}
\Delta M = k_B T_\kappa \ln 2 \,,
\end{equation}
which has the same functional form as Landauer’s principle, Eq.~(\ref{lds}). Accordingly, the mass loss \(\Delta M\) depends solely 
on the temperature. This explains why the red line in Fig.~2 shows that black holes of different masses exhibit the same \(\Delta M\), 
indicating that they have the same temperature.

The second aspect is that the black hole mass loss in the modified Kaniadakis framework, Eq.~(\ref{dmassk}), is always greater than or 
equal to that predicted by the standard Bekenstein--Hawking entropy, Eq.~(\ref{ldm}), as shown in Fig.~2. This suggests a shorter 
evaporation time in the modified Kaniadakis framework, which may be consistent with the non-observation of primordial black holes. 
For a more detailed discussion, see, for example, Ref.~\cite{cksy}.

\section{Conclusions}

We have proposed a modified Kaniadakis entropy, originally conceived with the aim of achieving thermodynamic stabilization of 
black holes \cite{aa}. This proposition is underpinned by equations connecting the two entropies, such as Eq. (\ref{kbh}), which establishes a 
relationship between the number of microstates $W$ and the BH entropy $S_{BH}$. Similar to the conventional Tsallis and Kaniadakis entropies, 
the modified Kaniadakis entropy offers a generalized framework for statistical mechanics through the introduction of a parameter $\kappa$.

Considering the MOND theory, which offers an alternative explanation for galaxy rotation curves compared to dark matter models, 
it represents a modification of Newton's second law. The theory introduces a function $\mu(x)$, exhibiting specific characteristics, 
to adjust the gravitational force in low-acceleration regimes and, in this study, we have obtained  a consistent interpolation function 
given by Eq. (\ref{inter}) which suggests that the effective force, Eq. (\ref{fmond2}), mimics MOND theory in certain aspects.

Regarding the Bekenstein bound conjecture, it posits a universal upper limit on the entropy of a confined quantum system. 
This conjecture has been demonstrated in various scenarios, including BH thermodynamics, where it imposes a constraint on the maximum entropy of a black hole system.
Analyzing the Bekenstein bound conjecture in the context of modified Kaniadakis entropy, Eq. (\ref{kbb2h}) suggests its fulfilment  
for typical values of $\kappa$, as illustrated by the ratio $R_\kappa$ in Figure 1. Particularly, when $\kappa = 0$, $R_\kappa$ 
equals 1, affirming the conjecture validity. Moreover, the fact that $R_\kappa$ increases with $\kappa$ indicates that the modified 
Kaniadakis entropy respects the Bekenstein bound conjecture.

Concerning the Landauer principle, we find that the mass loss of a black hole during Hawking radiation emission can be proportional
to its mass starting from a certain mass value onward when using the modified Kaniadakis entropy, as shown in Figure 2.
It is important to reiterate  that an important article \cite{bmenin} seeks to integrate Bekenstein's bound conjecture and 
Landauer's principle, presenting 
a unified framework for understanding the fundamental limits of entropy and energy in black hole systems.

Therefore, we can assert that
the exploration of a Kaniadakis-type statistics, alongside with its applications in various physical phenomena and its potential 
implications for BH thermodynamics, the Bekenstein bound conjecture, and the Landauer principle presents an interesting area of 
study at the intersection of statistical mechanics, gravitational phenomena and information theory.

\section*{AUTHORS’ DECLARATIONS}

\begin{itemize}
\item \textbf{Ethical conduct:} this manuscript complies with the ethical policies of the journal.
\item \textbf{Competing interests:} the authors have no competing interests that might influence the results
and/or discussion reported in this paper.
\item \textbf{Data availability:} all of the material is owned by the authors and/or no permissions have been
required. No data sets have been generated during the current study.
\item \textbf{Funding:} this work is partially supported by CAPES and FAPEMIG (Brazilian Research Agencies). 
Jorge Ananias Neto thanks CNPq (Conselho Nacional 
de Desenvolvimento Cient\'ifico e Tecnol\'ogico) for partial financial support, CNPq-PQ, 
Grant number 305984/2023-3.
\end{itemize}

\end{document}